\documentclass[
superscriptaddress,
twocolumn,
10pt,
prx,
aps,
floatfix,
nofootinbib,
sectionbib,
]{revtex4-2}

\input{glyphtounicode}
\usepackage[T1]{fontenc}
\usepackage[utf8]{inputenc}
\usepackage[english]{babel}
\babeladjust{autoload.bcp47 = on}
\usepackage{amsmath}  
\usepackage{amsfonts} 
\usepackage{amssymb}
\usepackage{graphicx} 
\usepackage[pdftex,bookmarks=true,bookmarksopen,bookmarksnumbered,colorlinks,linkcolor=blue,citecolor=blue,urlcolor=blue]{hyperref}
\graphicspath{ {./} }

\usepackage{tabularx}
\usepackage{array}
\usepackage[table]{xcolor}
\usepackage{xcolor}
\usepackage{braket}
\usepackage{booktabs}
\usepackage{multirow}
\usepackage{caption}
\usepackage{hyperref}
\usepackage{dsfont} 

\usepackage[resetlabels,labeled]{multibib}
\usepackage{comment}
\usepackage[separate-uncertainty=true]{siunitx}
\usepackage[shortlabels]{enumitem}

\usepackage{lipsum}

\newcites{Meth}{Methods References}
\newcites{Supp}{Supplementary References}

\newcommand{\aref}[1]{\hyperref[#1]{Appendix~\ref*{#1}}}
\DeclareCaptionLabelSeparator{line}{\hspace{2pt}\bf{|}\hspace{2pt}}

\DeclareMathOperator{\UU}{\ket{\uparrow\uparrow}}
\DeclareMathOperator{\DU}{\ket{\downarrow\uparrow}}
\DeclareMathOperator{\UD}{\ket{\uparrow\downarrow}}
\DeclareMathOperator{\DD}{\ket{\downarrow\downarrow}}

\newcommand{\nocontentsline}[3]{}
\newcommand{\tocless}[2]{\bgroup\let\addcontentsline=\nocontentsline#1{#2}\egroup}

\begin{document}

\captionsetup[figure]{name={\bf{Fig.}},labelsep=line,justification=centerlast,font=small}
\renewcommand{\equationautorefname}{Eq.}
\renewcommand{\figureautorefname}{Fig.}
\renewcommand*{\sectionautorefname}{Sec.}

\title{Precision high-speed quantum logic with holes on a natural silicon foundry platform}


\author{Isaac Vorreiter}
\email{i.vorreiter@unsw.edu.au}
\affiliation{School of Physics, University of New South Wales, Sydney, NSW 2052, Australia}

\author{Jonathan Y. Huang}
\affiliation{School of Physics, University of New South Wales, Sydney, NSW 2052, Australia}

\author{Scott D. Liles}
\affiliation{School of Physics, University of New South Wales, Sydney, NSW 2052, Australia}

\author{Joe Hillier}
\affiliation{School of Physics, University of New South Wales, Sydney, NSW 2052, Australia}

\author{Ruoyu Li}
\affiliation{IMEC, Leuven, Belgium}

\author{Bart Raes}
\affiliation{IMEC, Leuven, Belgium}

\author{Stefan Kubicek}
\affiliation{IMEC, Leuven, Belgium}

\author{Julien Jussot}
\affiliation{IMEC, Leuven, Belgium}

\author{Sofie Beyne}
\affiliation{IMEC, Leuven, Belgium}

\author{Clement Godfrin}
\affiliation{IMEC, Leuven, Belgium}

\author{Sugandha Sharma}
\affiliation{IMEC, Leuven, Belgium}

\author{Danny Wan}
\affiliation{IMEC, Leuven, Belgium}

\author{Nard Dumoulin Stuyck}
\affiliation{School of Electrical Engineering and Telecommunications, University of New South Wales, Sydney, NSW 2052, Australia}
\affiliation{Diraq Pty. Ltd., Sydney, NSW, Australia}

\author{Will Gilbert}
\affiliation{School of Electrical Engineering and Telecommunications, University of New South Wales, Sydney, NSW 2052, Australia}
\affiliation{Diraq Pty. Ltd., Sydney, NSW, Australia}

\author{Chih Hwan Yang}
\affiliation{School of Electrical Engineering and Telecommunications, University of New South Wales, Sydney, NSW 2052, Australia}
\affiliation{Diraq Pty. Ltd., Sydney, NSW, Australia}

\author{Andrew S. Dzurak}
\affiliation{School of Electrical Engineering and Telecommunications, University of New South Wales, Sydney, NSW 2052, Australia}
\affiliation{Diraq Pty. Ltd., Sydney, NSW, Australia}

\author{Kristiaan De Greve}
\affiliation{IMEC, Leuven, Belgium}
\affiliation{Department of Electrical Engineering (ESAT), KU Leuven, Leuven, Belgium}

\author{Alexander R. Hamilton}
\email{alex.hamilton@unsw.edu.au}
\affiliation{School of Physics, University of New South Wales, Sydney, NSW 2052, Australia}

\date{\today}

\begin{abstract}

\textbf{Silicon spin qubits in gate-defined quantum dots leverage established semiconductor infrastructure and offer a scalable path toward transformative quantum technologies. Holes spins in silicon offer compact all-electrical control, whilst retaining all the salient features of a quantum dot qubit architecture. However, silicon hole spin qubits are not as advanced as electrons, due to increased susceptibility to disorder and more complex spin physics. Here we demonstrate single-qubit gate fidelities up to 99.8~\% and a two-qubit gate quality factor of 240, indicating a physical fidelity limit of 99.7~\%. These results represent the highest performance reported in natural silicon to date, made possible by fast qubit control, exchange pulsing, and industrial-grade fabrication. Notably, we achieve these results in a near-identical device as used for highly reproducible, high-fidelity electron spin qubits. With isotopic purification and device-level optimisations in the future, our hole spin qubits are poised to unlock a new operation regime for quantum CMOS architectures.
}

\end{abstract}

\maketitle
\newpage

Among the diverse approaches to quantum computing, spin qubits in silicon are increasingly recognised as a competitive platform due to their compatibility with industrial semiconductor technology and potential for large-scale integration~\cite{veldhorst2017silicon,li2018a,gonzalez-zalba2021scaling,kunne2024the,siegel2024towards}. Hole spin qubits, which use the $j_z = \pm3/2$ ground state to encode the $\ket{0}$ and $\ket{1}$ qubit states, stand out due to their strong spin-orbit interaction (SOI), which unlocks ultrafast spin control via electric-dipole spin resonance (EDSR)~\cite{bulaev2007electric,crippa2018electrical,froning2021ultrafast,hendrickx2020fast,camenzind2022a,fang2023recent} and facilitates spin-photon coupling~\cite{yu2023strong}. Unlike electrons, the $p$-wave nature of holes dramatically suppresses unwanted nuclear hyperfine noise~\cite{testelin2009hole,degreve2011ultrafast,bosco2021fully}. Although holes confined in quantum dots have been realised in several different silicon MOS architectures~\cite{maurand2016cmos, camenzind2022a,geyer2024anisotropic}, the simultaneous fulfilment of all DiVincenzo criteria~\cite{loss1998quantum,divincenzo1997topics,divincenzo1998quantum,divincenzo2000the} remains to be demonstrated.

Utilising a $p$-type device fabricated in a modern, cutting-edge $\SI{300}{\milli\meter}$ CMOS process~\cite{li2020a}, we implement all essential components required for scaling up hole spin qubits -- single-shot readout, deterministic initialisation, and one- and two-qubit logic. Importantly, we are able to demonstrate high-fidelity one- and two-qubit gates in natural silicon with 4.7~\% $^{29}\mathrm{Si}$ spinful nuclei, without the need to isotopically purify the silicon substrate. We find the quality factors of single- and two-qubit gates corresponding to physical fidelity limits well above the surface code threshold~\cite{fowler2012surface}. We measure single-qubit fidelities that represent the highest level in natural silicon MOS quantum dots and perform an initial study on the physical properties of the qubits. These results are obtained in devices with near-identical geometries and fabrication procedures used in recent high-fidelity electron spin qubit measurements~\cite{steinacker2024300mm} (the main difference being the peripheral $p$-type contacts). This platform therefore enables the possibility of quantum CMOS chips that combine both electrons and holes.

\section{Device operation and readout}

\begin{figure*}[ht!]
    \includegraphics[width=\linewidth]{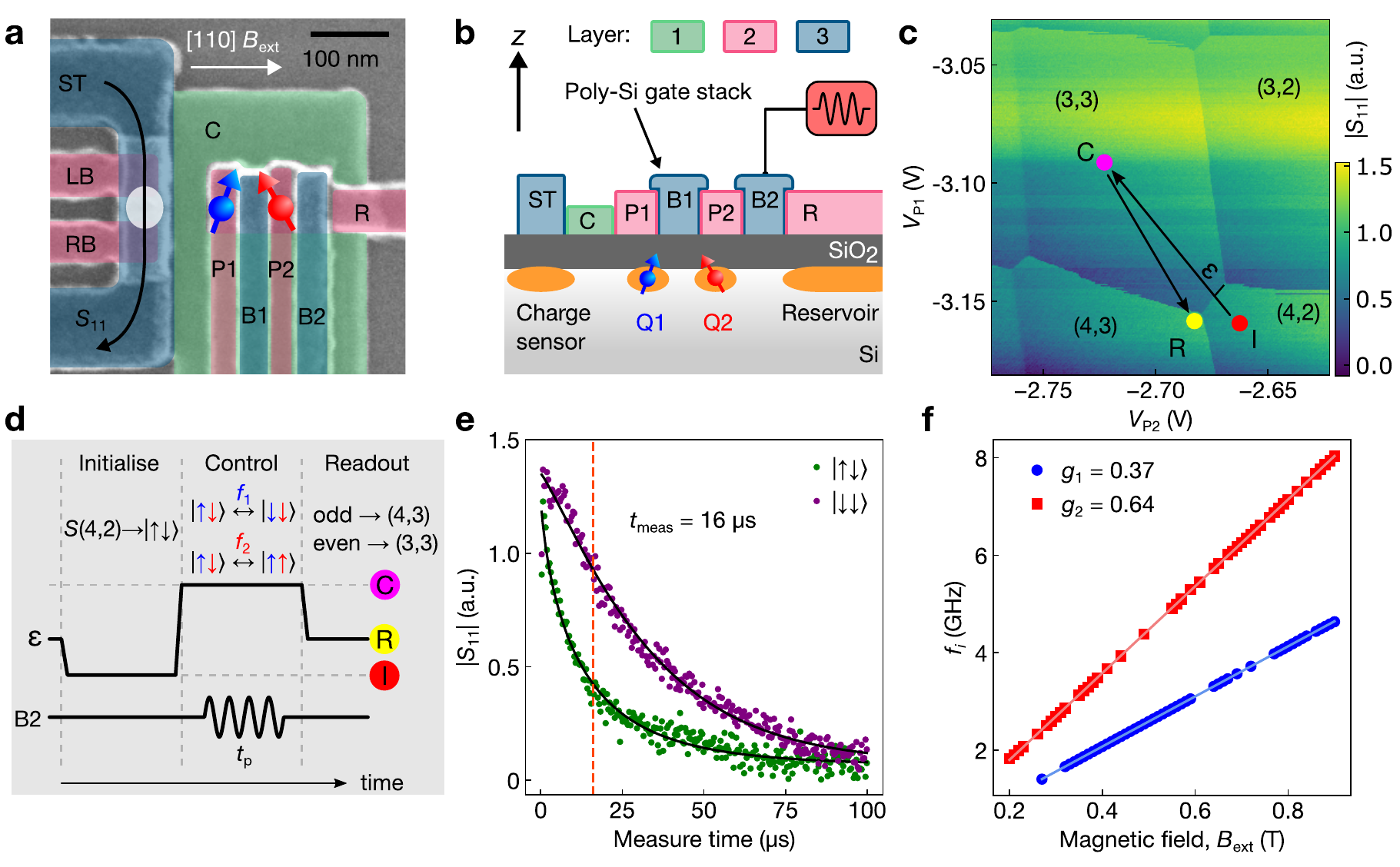}
    \caption{\textbf{Operation and readout of a hole-based two-qubit unit cell in natural silicon MOS.} 
    \textbf{(a)} False-coloured scanning electron microscope (SEM) image and \textbf{(b)} schematic of the device. Qubits Q1 and Q2 are formed in quantum dots defined by the C, B1, and B2 gates, and the hole occupation is controlled using the P1 and P2. Qubit control is performed using a microwave tone on B2, and single shot spin-to-charge readout is performed with radio frequency reflectometry on an adjacent sensor dot. 
    \textbf{(c)} Stability diagram at the (3,3)-(4,2) region where the qubits are operated, equivalent to a (1,1)-(2,0) transition. Single shot readout is performed via latched Pauli Spin Blockade.
    \textbf{(d)} Pulse scheme for single-spin manipulation, consisting of spin initialisation (I), control (C) and measure (M) steps. In the control step an $X(\theta)$ gate is created by applying a microwave pulse resonantly to one of the qubits to rotate the spin by angle $\theta$.
    \textbf{(e)} Charge sensor signal following the measurement scheme in (d) as a function of measure time for states in the $\ket{\uparrow\downarrow}$ and $\ket{\downarrow\downarrow}$ states. Readout fidelity is maximised when $t_\mathrm{meas}=\SI{16}{\micro\second}$.
    \textbf{(f)} Individual qubit addressability: qubit Larmor frequencies $f_i$ as a function of magnetic field $B_\mathrm{ext}$. Linear fits to $f_i = g_i\mu_\mathrm{B} B_\mathrm{ext}/h$ yield $g$-factors of $g_1 = 0.37$ and $g_2 = 0.64$. All measurements performed in a dilution refrigerator with a base temperature of \SI{10}{\milli\kelvin}.}
    \label{fig:main_fig_1}
\end{figure*}

Fig.~\ref{fig:main_fig_1}a shows a SEM of the device, and Fig.~\ref{fig:main_fig_1}b shows a corresponding schematic of the device cross-section. The device is fabricated on a $\SI{10}{\nano\meter}$ SiO\textsubscript{2} oxide on an undoped natural silicon substrate. The gates consist of a 3-layer polysilicon stack fabricated using electron beam lithography and subtractive patterning~\cite{elsayed2022low}.

The two quantum dots are defined under the plunger gates P1 and P2, which are used to control the dot occupation. The dots are confined laterally using the C gate and barrier gates B1 and B2, and the charge occupations are controlled using the plunger gates P1 and P2. A 2D reservoir of holes is formed under the R gate. The hole occupancy in each dot is measured with an adjacent single-hole transistor (SHT) integrated with radio-frequency (RF) reflectometry~\cite{angus2008a}, which provides a high readout bandwidth of $\sim\SI{2}{\mega\hertz}$.

To configure our hole quantum dots as spin qubits, we tune the double dot system into the weakly-coupled regime, where the tunnel coupling is much less than the thermal energy of the reservoir $t_\mathrm{c} \ll k_\mathrm{B}T$. We control the number of holes in each dot with P1 and P2, monitored with the charge sensor, as shown in Fig.~\ref{fig:main_fig_1}c. The periodic loading of each of the two dots, and the interdot coupling, leads to the characteristic `honeycomb’ structure of a double-dot charge stability diagram. In this way, we track the dot occupation down to the last hole in each dot, demonstrating full control over the absolute number of holes in each dot. Knowing the occupation is important, as it sets the nature of the ground state which strongly influences the properties of the qubit.

Numerical calculations of the electrostatic potential in these planar MOS devices shows that the holes are strongly confined against the heterointerface, into a `pancake'-like quantum dot with characteristic length scales of $\sim\SI{10}{\nano\meter}$ in the z-plane and $\sim\SI{50}{\nano\meter}$ in the xy-plane~\cite{liles2021electrical}. This confinement lifts the degeneracy of the light and heavy hole bands, with only heavy hole states states occupied, and the light hole states $\sim\SI{10}{\milli\electronvolt}$ higher in energy~\cite{marcellina2017spin}.

To operate the spins as qubits, we need the ability to initialise and measure the spin states. Fig.~\ref{fig:main_fig_1}c shows the stability diagram in the (3,3)-(4,2) regime, where $(N_1,N_2)$ denotes the hole occupation of the dots under P1 and P2, respectively. These hole occupations are chosen to allow readout via Pauli Spin Blockade (PSB)~\cite{seedhouse2021pauli,ono2002current,lai2011pauli}, since spin-wise, they map to an equivalent (1,1)-(2,0) regime~\cite{liles2018spin}. A single shot measurement is performed using the pulse sequence shown in Fig.~\ref{fig:main_fig_1}d, whose trajectory in the voltage space is shown in Fig.~\ref{fig:main_fig_1}c. The pulse sequence consists of three steps performing (I)nitialise -- (C)ontrol -- and (M)easure. Firstly, the spins are initialised at (I) in the (4,2) region as a singlet state, before being pulsed along the P2-P1 detuning axis $\varepsilon$ into (3,3) where they form a $\UD$ state. Secondly, single qubit control is performed within the (3,3) region at (C) using microwave pulses. Lastly, a projective spin readout is performed at the (M) point using PSB. Additionally, we perform a reference measurement immediately after the pulse sequence to subtract background noise from the result.

In holes, the readout visibility using PSB is reduced in the presence of a large $g$-factor difference $\Delta g$~\cite{seedhouse2021pauli}. To improve the visibility, we augment the spin-to-charge conversion with a latching step, where the singlet and triplet states map to different hole occupations~\cite{patrickcollard2018high,liles2024singlet}. In this regime, the $T_0(3,3)$ triplet relaxes to the $S(3,3)$ singlet within our readout integration time $t_\mathrm{meas}=\SI{16}{\micro\second}$, therefore the readout distinguishes between even and odd parity instead of singlet and triplet states~\cite{seedhouse2021pauli,nurizzo2023complete}. In this work, we use the convention of measuring the probability of the even parity spin-blocked states $P_\mathrm{even}$. Ultimately, using this combination of PSB and charge latching, we can threshold the measurement outcome to perform single-shot readout in $t_\mathrm{meas}=\SI{16}{\micro\second}$. Having demonstrated control of the absolute hole number in each dot, as well as fast single shot readout and qubit initialisation, we now move on to coherent qubit control.

\section{Single-qubit logic}

\begin{figure*}[ht!]
    \includegraphics[width=\linewidth]{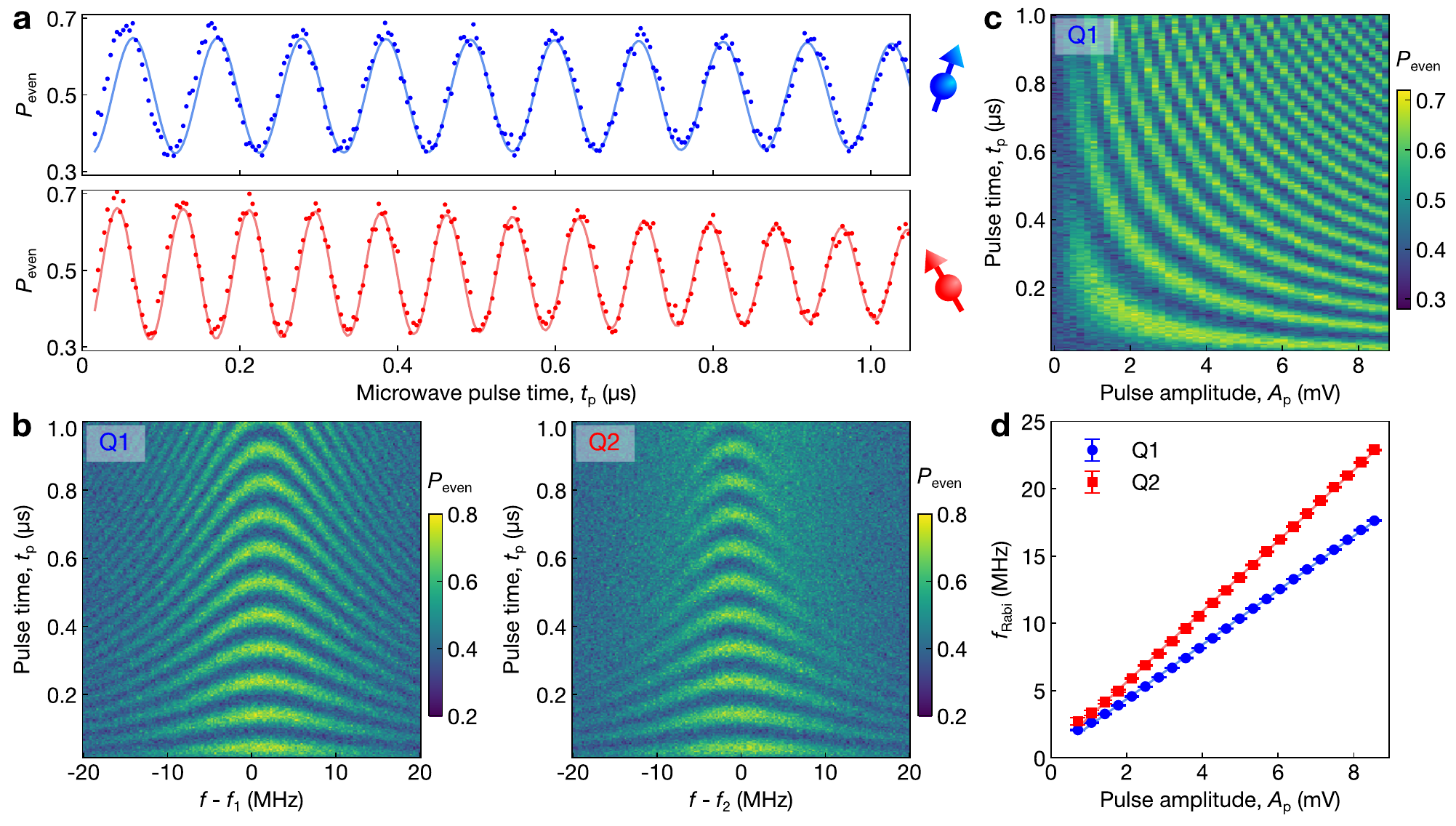}
    \caption{\textbf{Fast electrically driven single-qubit control for Q1 and Q2.} 
    \textbf{(a)} Spin state determined from the spin blocked probability as a function of resonantly-driven microwave pulse time for Q1 (blue) and Q2 (red). The solid lines are fits to a decaying sinusoid given by $P(t_\mathrm{p})= c_1 \cos (2 \pi f_\mathrm{Rabi} t_\mathrm{p} + \phi) \exp (-(t_\mathrm{p}/T_2^\mathrm{Rabi})^\alpha) + c_2$. From the fit we extract $T_2^\mathrm{Rabi}$ for each qubit as $\SI{3.95\pm0.13}{\micro\second}$ and $\SI{1.89\pm0.07}{\micro\second}$, with errors corresponding to the 95\% confidence interval of the fit. \textbf{(b)} Spin state probability as a function of microwave pulse time and frequency detuning producing a Rabi chevron for each qubit. 
    \textbf{(c,d)} Power dependence of the Rabi oscillations and Rabi frequency $f_\mathrm{Rabi}$, showing the expected linear dependence as a function of microwave pulse amplitude $A_\mathrm{p}$. All measurements taken at $B_\mathrm{ext} = \SI{0.83}{\tesla}$.}
    \label{fig:main_fig_2}
\end{figure*}

To rotate the individual spins between the $\ket{\uparrow}$ and $\ket{\downarrow}$ states, we exploit the strong SOI to drive spins electrically using a microwave pulse applied to the B2 gate~\cite{bulaev2007electric}. When the frequency of the microwave pulse matches the \textit{i}th-qubit's Larmor frequency $f_i = g_i\mu_\mathrm{B}B_\mathrm{ext}/h$, the spin is driven coherently via EDSR. In Fig.~\ref{fig:main_fig_1}f, we plot the measured Larmor frequency as a function of magnetic field $B_\mathrm{ext}$. From this, we find the $g$-factors to be $g_1=0.37$ and $g_2=0.64$ for Q1 and Q2, respectively. We account for this difference in $g$-factor from the observed Stark shift ($\mathrm{d}g/\mathrm{d}V\approx \SI{0.7}{\per\volt}$), and the $\sim\SI{0.45}{\volt}$ difference in plunger gate voltages.

We investigate the coherent manipulation of the qubit states by driving rotations between the $\ket{\uparrow}$ and $\ket{\downarrow}$ states with EDSR. We demonstrate spin control by varying the length of an on-resonance microwave pulse, $t_\mathrm{p}$, during the (C)ontrol step. The resultant Rabi oscillations are shown in Fig.~\ref{fig:main_fig_2}a for both qubits. The frequency and decay of the Rabi oscillations is extracted from a decaying sinusoid fit, yielding Rabi frequencies of $\SI{9}{\mega\hertz}$ and $\SI{12}{\mega\hertz}$ and decay times of $3.95\pm\SI{0.07}{\micro\second}$ and $1.89\pm\SI{0.03}{\micro\second}$ for Q1 and Q2, respectively. These values can be cast in the context of quantum information processing by using the gate quality factor $Q_\mathrm{gate} = f_\mathrm{Rabi}T_2^\mathrm{Rabi}$ -- the number of gates that can be coherently executed. Our results show quality factors of 36 and 23 for Q1 and Q2, which are reasonable in the context of spin qubits~\cite{stanoreview2022}. The gate quality factor may be improved by tuning the qubit into a `sweet spot' using magnetic field direction and gate voltage~\cite{venitucci2018electrical, wang2024electrical}.

In Fig.~\ref{fig:main_fig_2}b, we extend the Rabi experiment by varying the pulse length for different frequency detunings $\Delta f_i = f-f_{i}$ of the microwave pulse. This produces a Rabi chevron for both qubits.
In Fig.~\ref{fig:main_fig_2}c, we plot measured Rabi oscillations for different microwave pulse amplitudes $A_\mathrm{p}$. The extracted Rabi frequencies are plotted against $A_\mathrm{p}$ in Fig.~\ref{fig:main_fig_2}d, which displays a linear relationship that confirms $f_\mathrm{Rabi} \propto A_\mathrm{p}$. This leads to a maximum $f_\mathrm{Rabi}=\SI{23}{\mega\hertz}$, or equivalently, a minimum gate time of $t_{\pi/2}=\SI{10}{\nano\second}$. In our experiment, the speed is limited by the attenuation in the coaxial lines and the maximum amplitude we can apply using our microwave source, preventing us from seeing a non-linear driving regime. We predict that the maximum amplitude we could apply to the qubit is limited by the charging energy voltage $V_C\sim\SI{100}{\milli\volt}$, leading to a bound on the Rabi frequency of $\sim\SI{200}{\mega\hertz}$. In terms of driving speed, the measured hole spin qubits exceed electrons measured in near-identical devices~\cite{steinacker2024300mm} by a factor of $\approx 20$.

\section{Two-qubit logic}

\begin{figure*}[ht!]
    \includegraphics[width=0.8\linewidth]{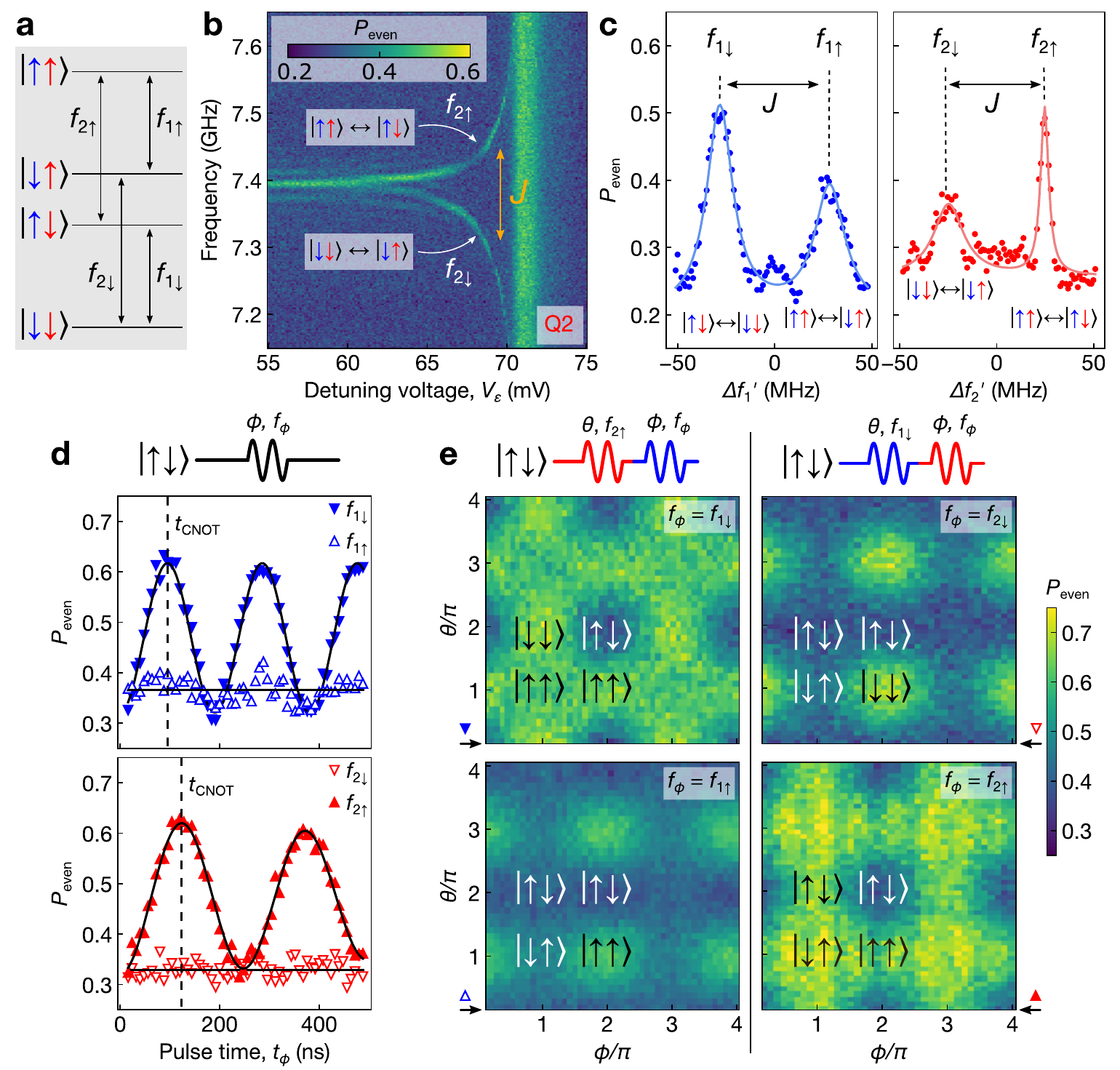}
    \caption{\textbf{Exchange interaction and controlled rotation (CROT) operation.} 
    \textbf{(a)} Energy level diagram of the four non-degenerate spin states in the presence of finite exchange $J$. 
    \textbf{(b)} EDSR spectrum as a function of P2-P1 detuning voltage $V_{\varepsilon}$. As detuning $\varepsilon$ approaches the anticrossing ($V_\varepsilon \rightarrow \SI{72}{\milli\volt}$), the exchange interaction splits the single qubit frequency into separate branches corresponding to the $\UD\leftrightarrow \UU$ and $\DU \leftrightarrow \DD$ transitions. 
    \textbf{(c)} EDSR spectrum vs frequency detuning taken at $V_\varepsilon = \SI{66}{\milli\volt}$ demonstrating an exchange splitting of $J = 53\pm\SI{4}{\mega\Hz}$ for both qubits. Fits to the data are made using a bimodal Lorentzian.
    \textbf{(d)} CROT: $P_\text{even}$ as a function of the qubit pulse time $t_\phi$ on Q1, demonstrating that the rotation of one qubit is conditional on the spin state of the other. Lower panel is same but for Q2. 
    \textbf{(e)} Full CROT control diagram: Same as for (d), but with an additional pulse of length $t_\theta$ and frequency $f_{1\downarrow}$ (left panels) or $f_{2\uparrow}$ (right panels). This demonstrates two qubit logic.}
    \label{fig:main_fig_3}
\end{figure*}

A common approach to implementing two-qubit logic with spins is through the exchange coupling $J$ of a double quantum dot~\cite{loss1998quantum,petta2005coherent}. $J$ depends strongly on the wavefunction overlap between spins, meaning that $J$ increases rapidly as the two spins are brought closer either by pulsing $V_{\varepsilon}$ near the (3,3)-(4,2) anticrossing, or by biasing B1 to be more accumulating. The increase in $J$ results in a splitting between the individual qubit resonant frequencies, as shown in Fig.~\ref{fig:main_fig_3}a. This indicates that the \textit{i}th-qubit can only be addressed at $f_{i,\sigma}$, which depends on the spin state $\sigma$ of the other qubit.

To measure the exchange splitting, we first initialise in the $\UD$ state, which is not an eigenstate of the system near the anticrossing. As $\varepsilon$ approaches the anticrossing, the qubit resonance is split into two given by $J$. This corresponds to transitions between the $\UD \leftrightarrow \UU$ and $\DU \leftrightarrow \DD$ states, as shown for Q2 in Fig.~\ref{fig:main_fig_3}b, where we plot the EDSR spectrum versus $V_{\varepsilon}$. Here we observe that $J$ gets larger as $V_\varepsilon \rightarrow \SI{72}{\milli\volt}$, corresponding to the (3,3)-(4,2) anticrossing. We can extend this experiment by comparing the EDSR spectrum at ($V_\varepsilon=\SI{66}{\milli\volt}$) as shown in Fig.~\ref{fig:main_fig_3}c. This clearly demonstrates that both qubit resonance frequencies are split by $J=\SI{53 \pm 4}{\mega\hertz}$.

We explore two implementations of two-qubit logic gates -- the driven controlled rotation (CROT) gate~\cite{zajac2018resonantly, huang2019fidelity,geyer2024anisotropic,noiri2022fast,hendrickx2020fast} and the decoupled controlled-phase (DCZ) gate~\cite{watson2018a,hendrickx2021a,xue2022quantum,tanttu2024assessment,huang2024high}. To enact the CROT, we initialise into the $\UD$ state and apply a microwave pulse of length $t_\mathrm{\phi}$ at one of the four exchange-split resonance frequencies $f_{i,\sigma}$, followed by a readout operation. The resulting spin-blocked probability is shown in Fig.~\ref{fig:main_fig_3}d. In the case where we drive on either $f_{1\downarrow}$ or $f_{2\uparrow}$, we see oscillations in the spin-blocked probability. However, if we drive  at $f_{1\uparrow}$ or $f_{2\downarrow}$, the qubits remain in a spin-unblocked state. This demonstrates rotations on each qubit, conditional on the state of the other qubit. In our experiment, a CROT-$\pi$ rotation can be performed in $t_\mathrm{CROT}=\SI{100}{\nano\second}$ $(\SI{125}{\nano\second})$ for Q1 (Q2). Together with the extracted $T_2^\text{CROT}$ times (Supplementary section \ref{supplementary:extended_crot}), we calculate quality factors ranging between 13 and 38.

In the driven CROT scheme, the choice of which of the four $f_{i,\sigma}$ frequencies to use is arbitrary, since a two-qubit gate set only requires one entangling gate. Hence, we can confirm that all four frequencies indeed allow for CROT. To achieve this, we extend on the initial experiment by first applying a pulse on the control qubit of length $t_\theta$. This control pulse is at $f_{1\downarrow}$ if the target is Q2 (left column), or $f_{2\uparrow}$ if the target is Q1 (right column). The resulting oscillation ‘checkerboard’ maps are shown in Fig.~\ref{fig:main_fig_3}e. For $t_\theta=0$, the horizontal linecuts of each sub-panel exactly match the plots in Fig.~\ref{fig:main_fig_3}d as indicted by arrow at the base of each sub-panel. When the length of the first pulse is timed to give an odd number of spin rotations ($\theta = n\pi$), we observe an ‘inversion’ of the control operation: oscillations now occur for $f_{1\uparrow}$ (bottom-left panel), whereas no oscillations occur for $f_{1\downarrow}$ (top-left panel) as they did for $\theta=0$. Similarly, oscillations now occur for $f_{2\downarrow}$ (top-right panel), whereas no oscillations occur for $f_{2\uparrow}$ (bottom-right panel). The control operation is then ‘restored’ when $\theta = 2n\pi$. This set of measurements demonstrates control of one qubit state conditional on another qubit state, and the canonical controlled-NOT (CNOT) gate can then be realised by combining the CROT-$\pi$ gate with a virtual $\sqrt{\mathrm{Z}}$-gate~\cite{huang2019fidelity}.


\begin{figure*}[ht!]
    \includegraphics[width=\linewidth]{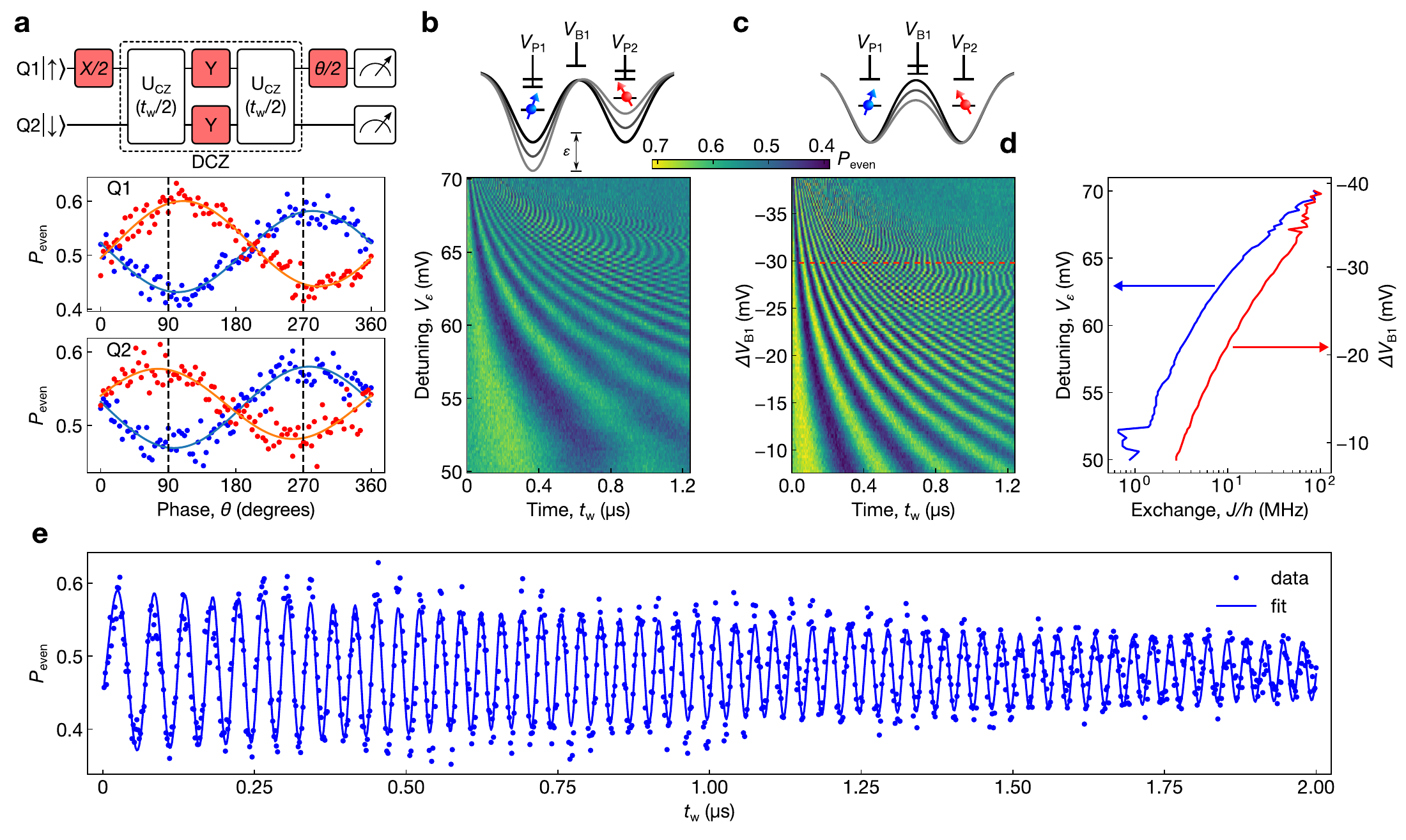}
    \caption{\textbf{Decoupled controlled phase (DCZ) operation.} 
    \textbf{(a)} Quantum circuit diagram used to perform the Ramsey-like calibration of the DCZ gate, with single-qubit driven operations indicated in red. Even-parity probability of the Ramsey-like sequence above outcome as a function of the phase of the second $\pi/2$ pulse, measured with exchange $J/h = \SI{20}{\mega\hertz}$.  For each measurement, we initialise the control qubit in its fiducial state (blue curve) or in its spin flipped state using a pi-pulse (red curve).
    \textbf{(b)} Measured DCZ oscillations as a function of the CZ operation time $t_w$ and the detuning pulse depth $V_\varepsilon$. 
    \textbf{(c)} Same as for (b), but with the exchange now controlled primarily by the interdot barrier gate B1.
    \textbf{(d)} Exchange $J$ extracted from the fast Fourier transforms of the DCZ oscillations in (b) and (c). Here we show that $J$ is tuneable over approximately 2 decades.
    \textbf{(e)} Linecut of DCZ exchange maps taken at $\Delta V_\text{B1}=\SI{-29.6}{\milli\volt}$. From the fit, we extract a  two-qubit gate quality factor of 240, corresponding to an upper limit of the fidelity of 99.7\%.
    }
    \label{fig:main_fig_4}
\end{figure*}
 
A simpler but faster implementation of two-qubit gates is the CZ family. When exchange is switched on, a qubit accumulates phase at different rates conditional on the state of the other qubit, with the difference given by $J$. To cancel out Stark shift on individual qubits and extend coherence, we incorporate a decoupling pulse to create DCZ operation, as illustrated in Fig.~\ref{fig:main_fig_4}~a. As a result of this pulse sequence, given a certain $J$ level, the phase of the target qubit oscillates at the same frequency with opposite phases when the control qubit is $\ket{\uparrow}$ and $\ket{\downarrow}$. A DCZ-$\pi$ operation can be implemented by calibrating the exchange wait time $t_\mathrm{w}$ such that the target qubit accumulates a phase of $\frac{\pi}{2}$ when the control qubit is $\ket{\uparrow}$ and $-\frac{\pi}{2}$ when the control qubit is $\ket{\downarrow}$. The canonical DCZ gate can then be formed by applying an unconditional phase correction of $\frac{\pi}{2}$ to both qubits after the exchange-echo-exchange process.

Fig.~\ref{fig:main_fig_4}b-c show DCZ oscillations as a function of $V_{\varepsilon}$ and the variation in B1 gate voltage $V_\mathrm{B1}$, respectively. As a result of the fast oscillations and the decoupling pulse, we acquire substantially increased quality factors. An example time trace is shown in Fig.~\ref{fig:main_fig_4}d, where $Q_\mathrm{gate}=240$ is observed. These quality factors, obtained from natural silicon, surpass those from the best academic-lab-built devices in isotopically purified silicon (from $\SI{800}{ppm}$ to $\SI{50}{ppm}$)~\cite{tanttu2024assessment,steinacker2025bell} and approach the level in state-of-the-art foundry devices with $\SI{400}{ppm}$ ${}^{29}$Si~\cite{steinacker2024300mm}.

\section{Coherence times and fidelities}

\begin{figure*}[t!h]
    \includegraphics[width=\linewidth]{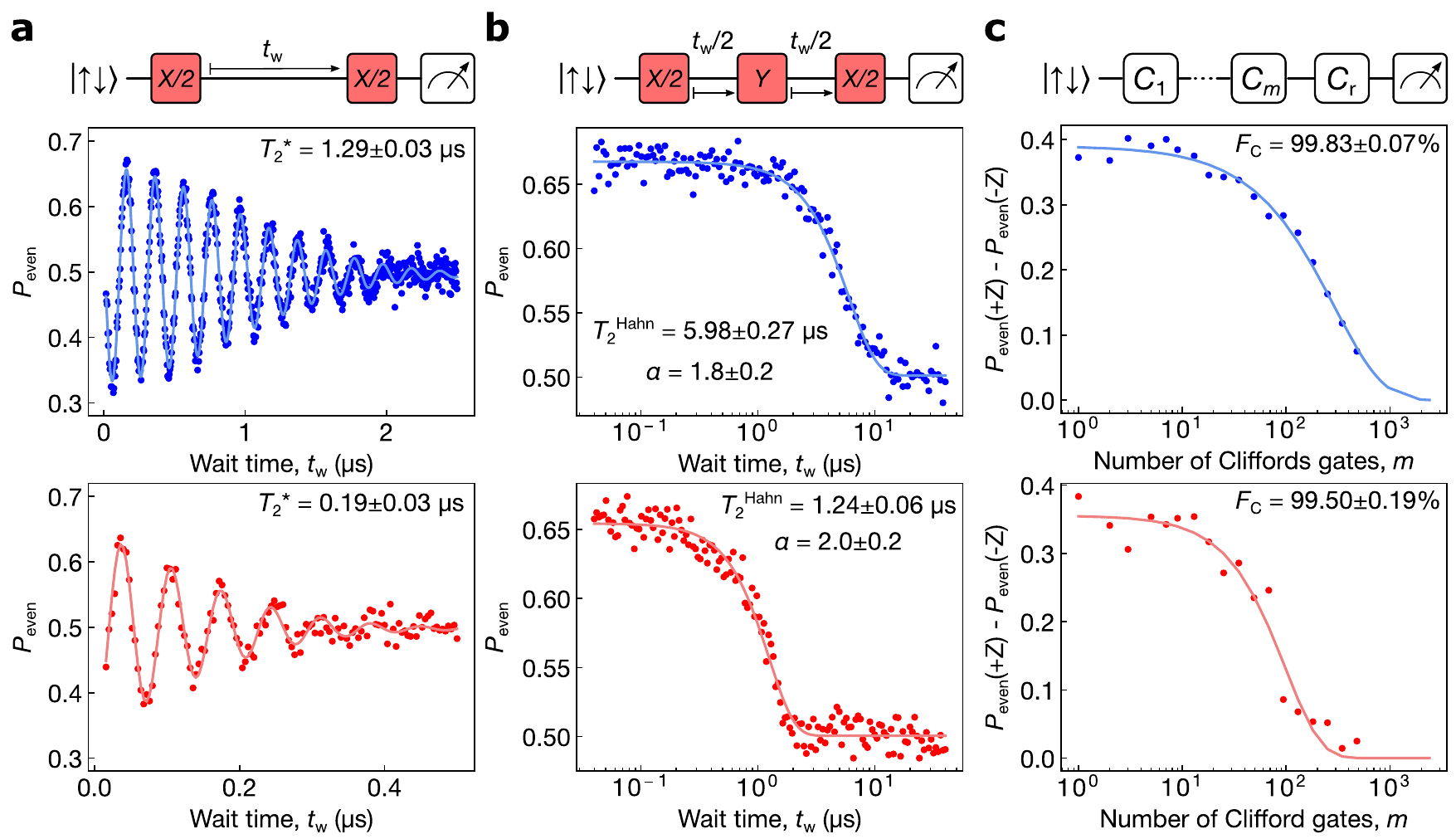}
    \caption{\textbf{Qubit performance.}
    \textbf{(a)} Single-qubit Clifford sequence for randomised benchmarking, with $m$ the number of Clifford gates. The decay of the sequence is fitted to $c_1 f^m+c_2$, from we extract a Clifford fidelities $F_\mathrm{C} = (1+f)/2$ of $99.83\pm\SI{0.07}{\percent}$ and $99.50\pm\SI{0.19}{\percent}$ for Q1 and Q2.
    \textbf{(b)} Free-induction decay / Ramsey sequence, showing resulting spin blocked probability as a function of wait time $t_\mathrm{w}$. The solid lines are fits to a decaying sinusoid $c_1 \cos (2\pi\Delta f t_\mathrm{w} + \phi) \exp (-(t_\mathrm{w}/T_2^*)^\alpha) + c_2$, from which we extract $T_2^* = 1.29\pm\SI{0.03}{\micro\second}$ and $0.19\pm\SI{0.03}{\micro\second}$ for Q1 and Q2.
    \textbf{(c)} Spin refocussing with a Hahn echo. By interleaving a $Y(\pi)$ gate in the Ramsey sequence, we refocus the spin dephasing and thereby extend the spin coherence time. Solid lines show fits to $c_1 \exp (-(t_\mathrm{w}/T_2^\mathrm{Hahn})^\alpha) + c_2$, yielding $T_2^\mathrm{Hahn} = 5.98\pm\SI{0.27}{\micro\second}$ and $1.24\pm\SI{0.06}{\micro\second}$.
    All experiments performed at $B_\mathrm{ext} = \SI{0.83}{\tesla}$. The errors correspond to the 95\% confidence interval of the fitting.}
    \label{fig:main_fig_5}
\end{figure*}

Qubit coherence is characterised by the $T_2^*$, $T_2^\mathrm{Hahn}$, and $T_2^\mathrm{CPMG}$ times, which reflect the dephasing from low, intermediate and high frequency noise sources. We evaluate the $T_2^*$ time through a Ramsey free-induction decay experiment. Here, we apply two $\mathrm{X_{\pi/2}}$ pulses with frequency detuning $\Delta\omega$, separated by a wait time $t_\mathrm{w}$. During the wait time, the qubit is exposed to low-frequency longitudinal noise, which causes the qubit to dephase. As shown in Fig.~\ref{fig:main_fig_5}a, this measurement produces oscillations with frequency $\Delta f$ in the spin probability modulated by a decay envelope. From this decay we extract coherence times of $T_2^* = 1.24\pm\SI{0.11}{\micro\second}$ and $0.19\pm\SI{0.02}{\micro\second}$ for Q1 and Q2. We can decouple the qubit from this low-frequency noise via a Hahn echo sequence, thereby increasing the coherence time of the qubit. This is achieved by interleaving an additional $Y(\pi)$ pulse into the Ramsey sequence, and the resulting spin echo is shown in Fig.~\ref{fig:main_fig_5}b. Here the spin probability of each qubit decays in characteristic time $T_2^\mathrm{Hahn} = 5.79\pm\SI{0.38}{\micro\second}$ and $1.25\pm\SI{0.06}{\micro\second}$ for Q1 and Q2, respectively. This confirms that refocussing extends the coherence time of both qubits.


One common source of dephasing for spin qubits is nuclear hyperfine noise. We estimate the dephasing from hyperfine noise by calculating the Overhauser field of the residual $^{29}\mathrm{Si}$ nuclei in natural silicon~\cite{philippopoulos2020first}. This yields an estimate of $\approx\SI{1.5}{\micro\second}$, which is of the same order of magnitude as our measured result for Q1. This finding is consistent with recent results in silicon hole spins~\cite{camenzind2022a, piot2022a, liles2024singlet} which indicates that hyperfine noise likely contributes significantly to the total dephasing rate. Furthermore, this dephasing rate is of similar order of magnitude to electrons in natural silicon~\cite{wang2024pursuing}, but a factor of $\approx10-100$ times lower than for electrons in 400 ppm isotopically enriched silicon~\cite{veldhorst2014addressable, steinacker2024300mm}. Using the same estimation as above, we would expect $T_2^*\approx\SI{10}{\micro\second}$ for 400 ppm isotopically enriched silicon. This would indicate that hyperfine noise may be further suppressed by isotopic purification of the silicon substrate, providing a simple path to significantly improving the hole qubit performance.

The second common source of dephasing for spin qubits is charge noise. This is typically present in the form of both 1/f noise and white noise. The ratio between the $T_2^*$ and the $T_2^\mathrm{Hahn}$, $T_2^\mathrm{CPMG}$ times is a proxy for determining the relative contribution of low-frequency noise to the overall dephasing time. In our study, we see $T_2^\mathrm{Hahn}/T_2^*$ $\approx$ 6 for both qubits. By comparison, for electron spins in 400 ppm silicon, the ratio of $T_2^\mathrm{Hahn}/T_2^*$ is $\approx$ 10~\cite{veldhorst2014addressable}, indicating that high-frequency noise contributes more to hole spins than for electron spins. Furthermore, we can get an indication of the noise colour at higher frequencies from the trend in the CPMG measurements (Supplementary section \ref{supplementary:cmpg_analysis}). From this analysis, we extract $\beta \approx 0.65$ which represents the scaling exponent of a power-law noise spectrum, namely $S_\varepsilon(f) \propto f^{-\beta}$.  Comparing this value of $\beta$ to the exponent of the measured low-frequency SHT charge noise spectrum, $\beta_\text{SHT}$=1.33, we see that it differs significantly (Supplementary section \ref{supplementary:charge_noise}). This discrepancy has been observed in other hole spin qubit systems~\cite{camenzind2022a, piot2022a}, and indicates the noise spectrum is closer to a white noise spectrum ($\beta$=0) at higher frequencies.

Qubit performance can be described more universally in terms of its fidelity, which evaluates the span of operations that are needed to perform two-axis control around the Bloch sphere. We evaluate the single qubit gate fidelity by performing standard randomised benchmarking (RB)~\cite{knill2008randomized, magesan2011scalable}, which is insensitive to state preparation and measurement (SPAM) errors. This technique evaluates the fidelity of all gates from the Clifford group, which quantifies the rates of errors in the gate set.

To benchmark single-qubit gates, we generate the Clifford group using X/2 and Z/2 gates as primitives, with the Z/2 gate implemented as a virtual gate using phase updates of the microwave signal. The Clifford measurement versus the number of Clifford gates $m$ is plotted in Fig.~\ref{fig:main_fig_5}c, showing a decay in probability as the number of Clifford gates increases. We perform each Clifford gate sequence with even and odd measurement projections, and subtract the resulting probabilities from each other to obtain the difference $P_\text{even}(+\mathrm{Z})-P_\text{even}(-\mathrm{Z})$ between the two final states. We fit the decay to the equation $c_1 f^{m}+c_2$ from which we extract the Clifford fidelity $F_\mathrm{C} = (1+f)/2$. We find a single qubit fidelities of $99.83\pm\SI{0.07}{\percent}$ and $99.50\pm\SI{0.19}{\percent}$ for Q1 and Q2. These fidelities exceed the fault-tolerant threshold of $\sim\SI{99.4}{\percent}$~\cite{fowler2012surface} and are amongst the highest values achieved in silicon-based hole spin qubits~\cite{stanoreview2022}, and in natural silicon in general. The factor of $\approx 3$ difference in infidelities between Q1 and Q2 can likely be attributed to their differences in Rabi (factor of $\approx 2$) and dephasing times (factor of $\approx 7$).

For two-qubit gates, we estimate the physical upper limit to the fidelity from the two-qubit $Q$-factors. We evaluate the fidelity $F_\text{2Q}$ from $F_\text{2Q}=3(1+f)/4$, where $f$ is extracted from fitting the decay function $c_1f^{m}+c_2$ to the randomised benchmarking experiment~\cite{knill2008randomized}. From the extracted $Q^\text{CROT}$ values, we obtain $F_\mathrm{CROT}$ ranging between $\SI{97}{\percent}$ and $\SI{99}{\percent}$. More impressively, the estimated upper limit for $F_\mathrm{DCZ}$ is $\SI{99.7}{\percent}$, considerably higher than the fault-tolerance threshold~\cite{fowler2012surface}. This leaves ample scope for further techniques -- such as substrate purification and advanced gate calibration -- to achieve practical gate fidelities above the threshold, as measured through GST in Ref.~\cite{steinacker2024300mm}.

Notably, our coherence times and gate performance metrics could be improved by optimising the magnetic field direction and gate biases, as holes in silicon exhibit both highly anisotropic and voltage-sensitive qubit properties~\cite{liles2021electrical,crippa2018electrical, piot2022a, carballido2024compromise}. Additionally, we also measure the qubit relaxation time $T_1 = \SI{255\pm60}{\micro\second}$ (Supplementary section \ref{supplementary:T1_measurement}), indicating that the dephasing time, rather than the relaxation time, is the limiting factor for these qubits. Follow-up studies could also explore the impact of gate-induced strain and device geometry on qubit properties, to provide insights into the variability of qubit performance as the system scales to larger numbers.

\section{Outlook}

Our work shows the feasibility of hole spin qubits in planar silicon MOS and introduces a valuable new addition to the toolbox of silicon spin-based quantum computing. It verifies that hole spin qubits in planar silicon are directly compatible with their electron spin analogues and can even be operated in identical device structures with similar control schemes. Furthermore, hole spin qubits can operate with fast and high-fidelity quantum gates. They also offer low-power all-electrical control and strong spin-photon coupling~\cite{yu2023strong}, providing key benefits to complement electron spin qubits that could be fabricated and operated on the same silicon CMOS chip.

The superior precision and quality of our quantum gates in natural silicon are surprising to witness yet physically grounded -- it is a proof-of-principle result of high-speed operation and the reduced hyperfine coupling of hole spin qubits to nuclear spins compared to electron spin qubits~\cite{testelin2009hole,degreve2011ultrafast,bosco2021fully}. As this coupling is still non-zero and tunable~\cite{bosco2021fully,fischer2010hybridization}, there is room in the future to further improve performance by isotopic purification and optimisations of quantum dot confinement and magnetic field orientation.



\section*{Methods}
\setcounter{subsection}{0}

\subsection{Experimental setup}\label{methods:experimental_setup}

The device is measured in a Bluefors LD250 dilution refrigerator with a base temperature of $\SI{10}{\milli\kelvin}$ and hole temperature of $\SI{130}{\milli\kelvin}$. An external fixed magnetic field is supplied by an American Magnetics AMI430 magnet points in the [110] direction of the silicon lattice. Low-frequency voltages are supplied with a Q-Devil QDAC-I, and the low-frequency lines pass through a Q-Devil $\SI{50}{\kilo\hertz}$ cold filter. Fast voltage pulses are generated with a Quantum Machines OPX+ and are combined with low-frequency biases using an RC bias tee with a crossover frequency of $\SI{145}{\hertz}$. Microwave pulses are generated by a Rhode \& Schwarz SGS100A Vector Signal Generator, using IQ upconversion. The in-phase and quadrature (I/Q) waveforms are generated by the OPX+. For the two-qubit control, a second SGS100A is used and the two microwave pulses are combined using a power combiner. Microwave pulses are applied to the device via grapho-coax, with additional $\SI{30}{\decibel}$ of in-line attenuation applied using discrete attenuators at different temperature stages within the fridge. Reflectometry is implemented by connecting the SHT ohmic to an external inductor $L=\SI{1200}{\nano\henry}$, which forms a tank circuit with the parasitic capacitance $C_\mathrm{p}$. This creates an impedance match at a resonant frequency of $f_0=\SI{156}{\mega\hertz}$. The reflectometry RF chain is configured with the incident signal delivered through $\SI{70}{\decibel}$ of in-line attenuation and a $\SI{12}{\decibel}$ directional coupler. The reflected signal is amplified at \SI{4}{\kelvin} using a Cosmic Microwave CITLF2 low-noise amplifier, followed at 300K by two $\SI{20}{\decibel}$ ZFL-1000LN+ low-noise amplifiers. A diagram of the measurement setup is shown in Supplementary section \ref{supplementary:experimental_setup}.




\section*{Data availability}
The datasets generated and/or analysed during this study are available from the corresponding authors on reasonable request.

\section*{Acknowledgements}
We thank L. Camenzind, A. Saraiva, A. Laucht, A. Chatterjee and W. A. Coish for helpful discussions. 
We acknowledge support from the Australian Research Council through grant LP200100019 and the US Army Research Office through W911NF-23-10092. This work was performed as part of IMEC’s Industrial Affiliation Program (IIAP) on Quantum Computing. I.V. acknowledges support from the Sydney Quantum Academy, A.R.H. acknowledges ARC Fellowship IL230100072 co-funded by Diraq Pty Ltd, and A.S.D. acknowledges ARC Fellowship FL190100167.

\section*{Author contributions}
I.V., J.Y.H., and J.H. performed the experiments.
I.V., J.Y.H., S.D.L., J.H., and A.R.H. contributed to discussions of the experimental results. 
I.V., J.Y.H., N.D.S., W.G., and C.H.Y. contributed to the experimental control software. 
I.V produced the figures for the manuscript, with input from all co-authors. The IMEC team developed
the \SI{300}{\milli\meter} spin qubit process, fabricated the device, and performed an initial electrical device screening at wafer-scale.
I.V, J.Y.H., and A.R.H. authored the manuscript, with reviews from all co-authors.

\section*{Corresponding authors}
Correspondence to I.V., or A.R.H.

\section*{Competing interests}
A. S. D. is CEO and a director of Diraq Pty Ltd. N. D. S, W.G., C. H. Y., and A. S. D. declare equity interest in Diraq. Other authors declare no competing interest.










\captionsetup[figure]{name={\bf{Fig.}},labelsep=line,justification=centerlast,font=small}
\renewcommand{\equationautorefname}{Eq.}
\renewcommand{\figureautorefname}{Fig.}
\renewcommand*{\sectionautorefname}{Sec.}
\renewcommand{\thesubsection}{S\arabic{subsection}}
\renewcommand{\thesection}{} 


\clearpage

\maketitle

\onecolumngrid

\section*{Supplementary information}

\setcounter{subsection}{0}
\setcounter{figure}{0}
\captionsetup[figure]{name={\bf{Supplementary  Fig.}}, labelsep=line, justification=centerlast, font=small}
\setcounter{table}{0}




\subsection{Experimental setup}\label{supplementary:experimental_setup}


\begin{figure*}[!ht] 
    \centering
    \includegraphics[width=0.65\linewidth]{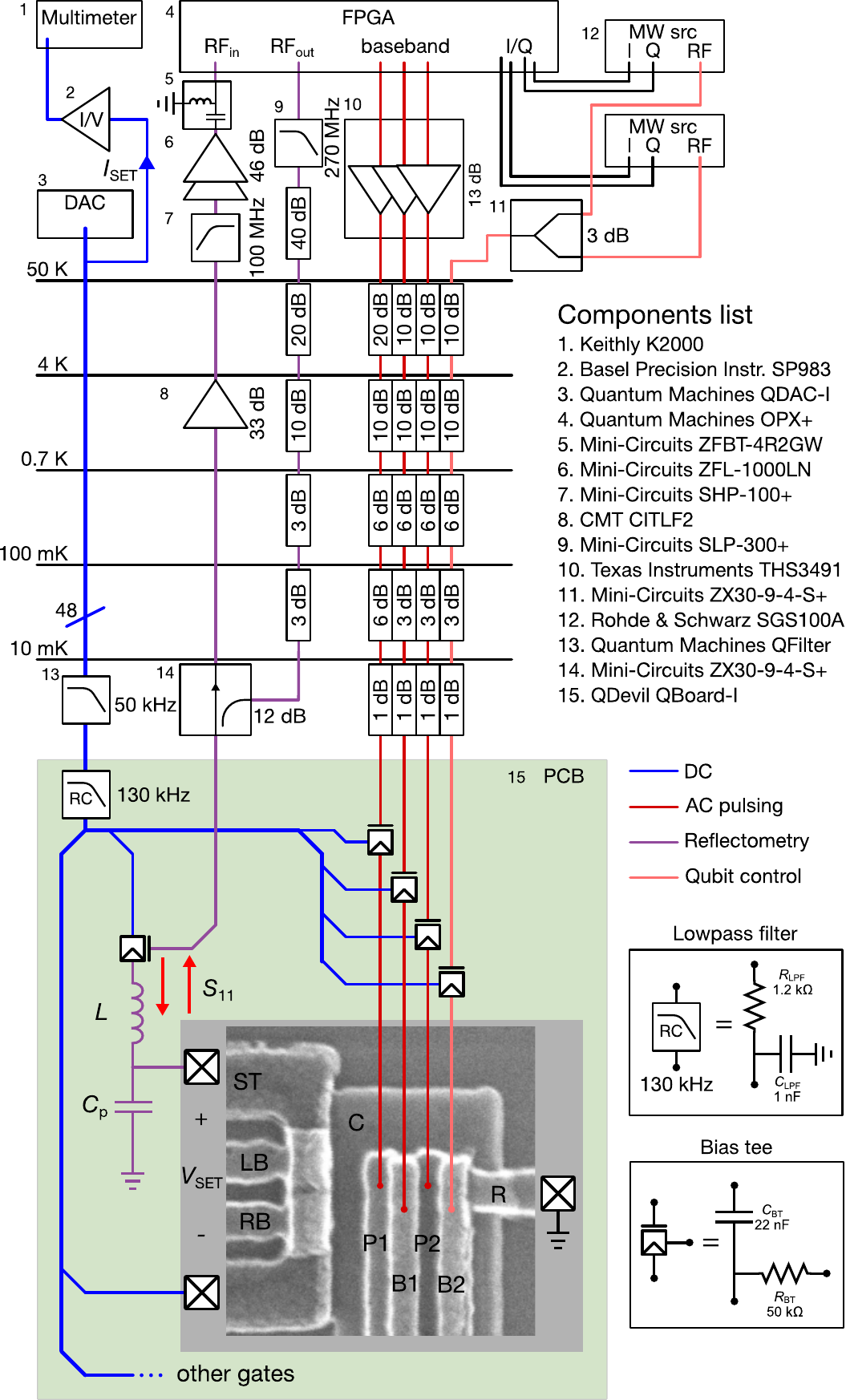}
    \caption{\textbf{Experimental setup.}
    }
    \label{fig:experimental_setup}
\end{figure*}

\clearpage

\subsection{Single hole transistor charge noise}\label{supplementary:charge_noise}

\begin{figure*}[!ht] 
    \centering
    \includegraphics[width=0.7\linewidth]{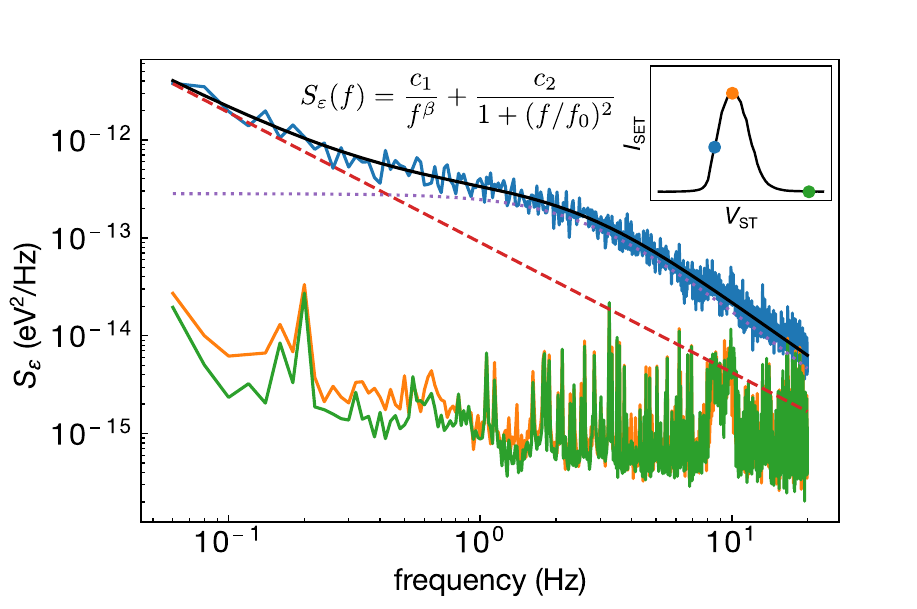}
    \caption{\textbf{Power spectral density of the charge noise of a single hole transistor.}
    Charge noise power spectrum as measured through the single hole transistor. The current trace is converted to a charge power spectral density $S_\varepsilon$. Charge noise amplitude at $\SI{1}{\hertz}$ is $\sqrt{S_\varepsilon}=\SI{0.63}{\micro\electronvolt\per\hertz\tothe{1/2}}$. The spectrum follows a composite power-law $c_1/f^\beta$ and Lorentzian $c_2/(1+(f/f_0)^2)$. From the fit (black), we extract a fit to the power law with an exponent of $\beta_\text{SHT}=\beta = 1.33\pm0.05$, and a fit to the Lorentzian with $f_0 = \SI{2.6 \pm 0.2}{\hertz}$. The individual power-law and Lorentzian fits are indicated by the dashed and dotted lines, respectively. Inset shows bias points on the SHT Coulomb oscillation where each spectrum was measured.
    }
    \label{fig:set_charge_noise}
\end{figure*}

\clearpage

\subsection{Extended charge stability diagram}\label{supplementary:stability_diagram}

\begin{figure*}[!ht] 
    \centering
    \includegraphics[width=0.5\linewidth]{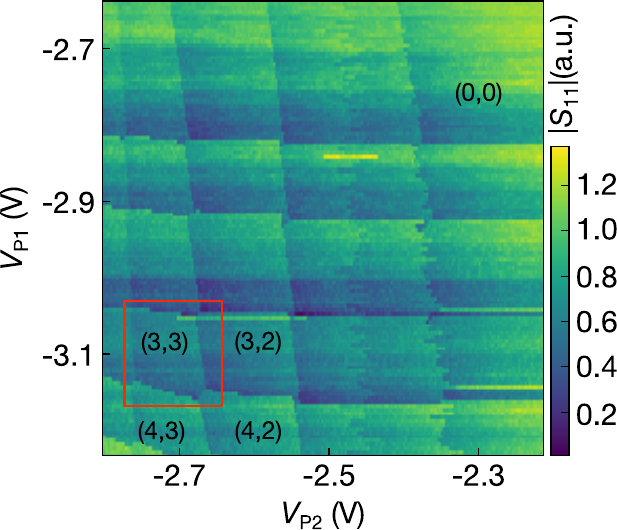}
    \caption{\textbf{Stability diagram showing control of the hole occupation in both dots down to the last hole.} Full stability diagram of the device down to the last hole. The colour axis is given by $|S_{11}|$, which is proportional to the conductance through the SHT charge sensor. The (3,3)-(4,2) operating region depicted in the main text is indicated by the red square.
    }
    \label{fig:csd_large}
\end{figure*}

\clearpage

\subsection{Dot lever arms and hole temperature}\label{supplementary:lever_arms}

\begin{figure*}[!ht] 
    \centering
    \includegraphics[width=\linewidth]{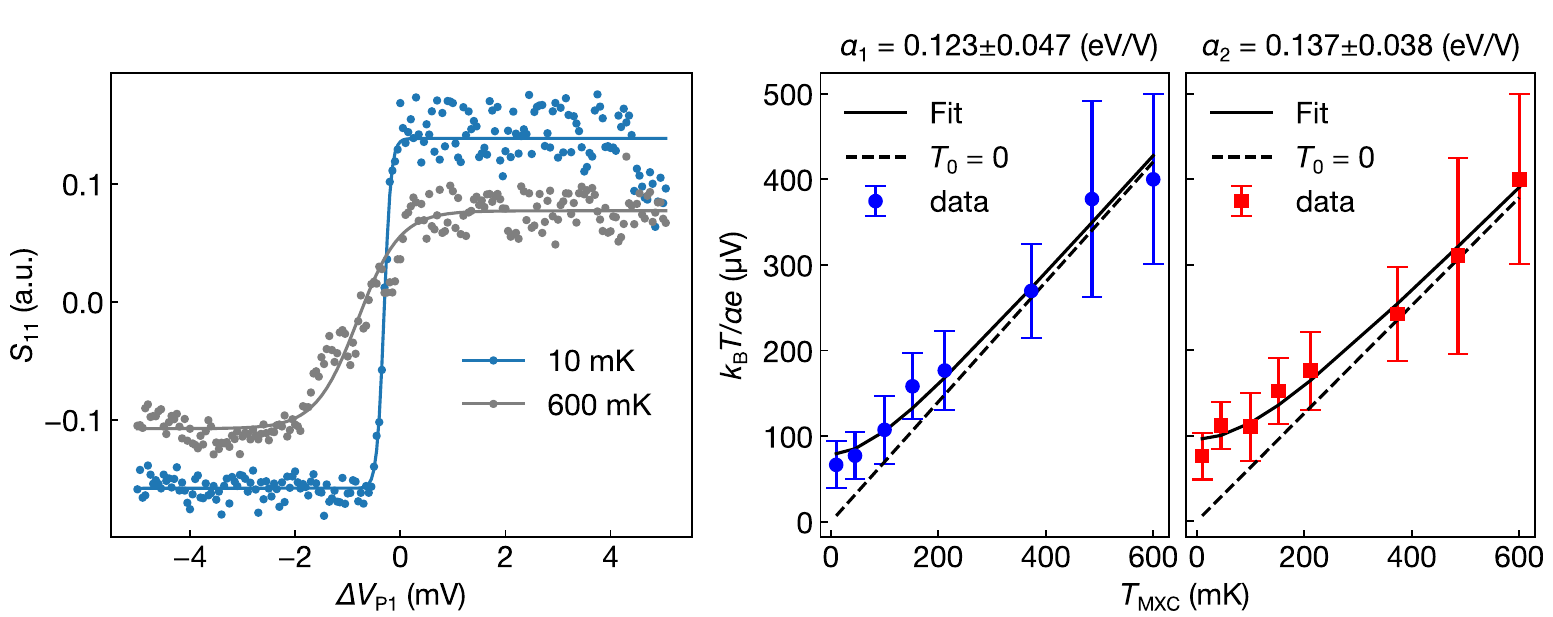}
    \caption{\textbf{Lever arm and hole temperature calculation.}
    \textbf{(a)} SHT conductance measured by sweeping the P1 gate voltage $V_\text{P1}$ at fridge temperatures of 15mK and 620mK. We repeat this measurement for the P2 gate voltage to extract both lever arms.
    \textbf{(b)} Plots of $k_BT/\alpha e$ resulting from the fits of the curves in (a) as a function of fridge mixing chamber temperatures $T_\text{MXC}$. From the fits we extract lever arms of $\alpha_1 = \SI{0.123\pm0.047}{\electronvolt\per\volt}$ and $\alpha_2 = \SI{0.137\pm0.038}{\electronvolt\per\volt}$ of the P1 and P2 plunger gates. The dotted lines show the inferred fit for the cases when $T_0=0$. Error bars are given from the 95\% confidence interval to the fits of the datasets in (a).
    }
    \label{fig:lever_arm}
\end{figure*}

In order to determine the charging energies, we need to be able to convert between the charging voltages (period of the stability diagram for $V_\text{P1}$ and $V_\text{P2}$) to an energy. For quantum dot systems, this conversion factor is known as the lever arm $\alpha$. Hence, we also need to ascertain the lever arms of P1 ($\alpha_1$) and P2 ($\alpha_2$) to their respective quantum dots in order to find the charging energy.

To find the lever arms, we measure the thermal broadening of the leads \cite{liles2018spin}. In Supplementary Fig.~\ref{fig:lever_arm}a we show the SHT reflectometry signal, $S_{11}$, for two different fridge mixing chamber temperatures $T_\text{MXC}$, measured across an $(N_1,N_2)\leftrightarrow (N_1+1,N_2)$ charge transition. The width of the transition is sensitive to the thermal broadening of the reservoir distribution of the hole states. We fit $S_{11}$ to the function 

$$S_{11}\propto\frac{1}{1+\exp\big(\frac{\Delta V_{\text{P1}}\alpha_i e}{k_BT}\big)} = \frac{1}{1+\exp\big( \frac{\Delta V_{\text{P1}}}{c_i}\big)}$$

where $\Delta V_\text{P1} = V_\text{P1} - V_0$, and $V_0$, $c_i=k_BT/\alpha_ie$ are fitting parameters. The smooth lines in Supplementary Fig.~\ref{fig:lever_arm}a show the fit to the raw data. We repeat these measurements for the $(N_1,N_2)\leftrightarrow (N_1,N_2+1)$ charge transition to extract the lever arm for P2 (not shown). From these fittings for P1 and P2, we plot $c_i=k_BT/\alpha_ie$ as a function of $T_\text{MXC}$ in Supplementary Fig.~\ref{fig:lever_arm}b for both P1 and P2 measurements. We then fit this data  to the function

$$y = \frac{k_BT}{\alpha_i e} = \frac{k_B}{\alpha_i e} \sqrt{T_\text{MXC}^2 + T_0^2} $$

where $\alpha_i$ and $T_0$ are fitting parameters. Here, $\alpha_i$ represents the $\text{P}_i$ gate lever arm, and $T_0$ represents the hole temperature. From the fits we extract lever arms of $\alpha_1 = \SI{0.123\pm0.047}{\electronvolt\per\volt}$ and $\alpha_2 = \SI{0.137\pm0.038}{\electronvolt\per\volt}$ of the P1 and P2 plunger gates. Additionally, we find an average hole temperature of $T_0=\SI{130\pm20}{\milli\kelvin}$. The measured values of the lever arm reflect typical values in silicon MOS quantum dots (typically $0.05$ to $\SI{0.2}{\electronvolt\per\volt}$).

\clearpage

\subsection{Extended single qubit Rabi oscillations}\label{supplementary:extended_rabi}

\begin{figure*}[!ht] 
    \centering
    \includegraphics[width=0.6\linewidth]{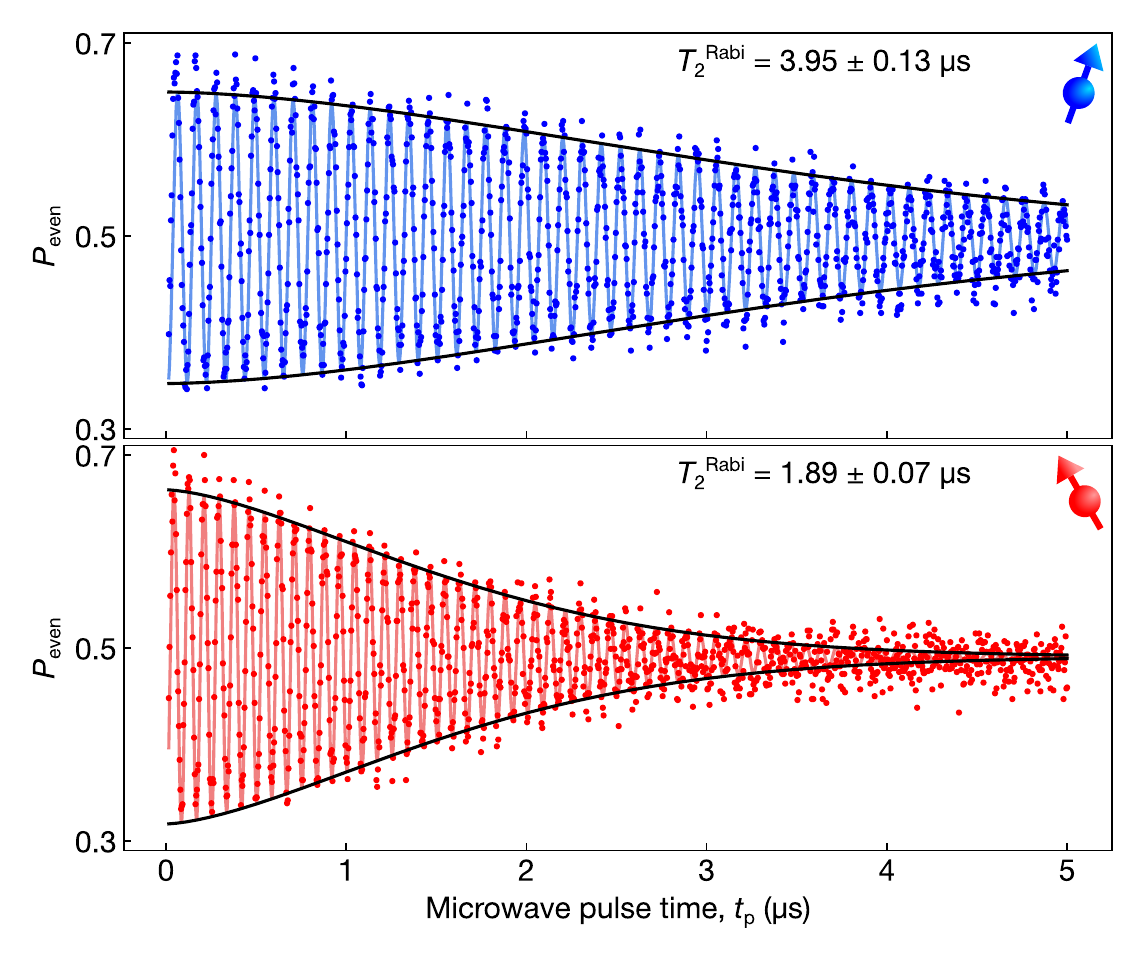}
    \caption{\textbf{Extended single qubit Rabi oscillations.} Rabi oscillations for both Q1 and Q2 showing the decay of the oscillations past the $1/e$ point.
    }
    \label{fig:rabi_extended}
\end{figure*}

\clearpage

\subsection{Extended CROT oscillations}\label{supplementary:extended_crot}

\begin{figure*}[!ht] 
    \centering
    \includegraphics[width=0.6\linewidth]{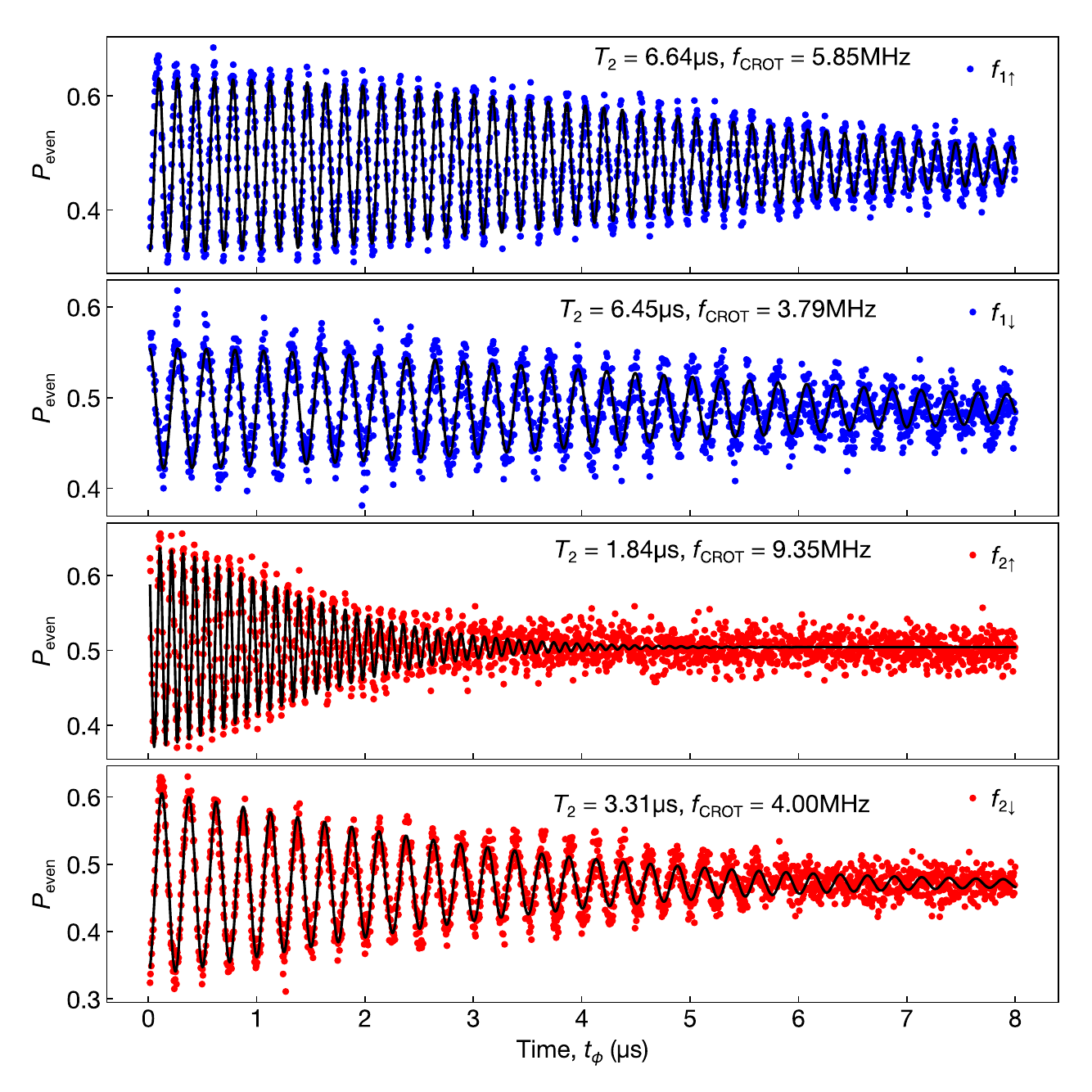}
    \caption{\textbf{Extended CROT oscillations.} CROT oscillations for all four exchange-split EDSR branches branches with the control `enabled'.
    }
    \label{fig:crot_extended}
\end{figure*}

\clearpage

\subsection{Dynamical decoupling and noise analysis}\label{supplementary:cmpg_analysis}

\begin{figure*}[!ht] 
    \centering
    \includegraphics[width=0.8\linewidth]{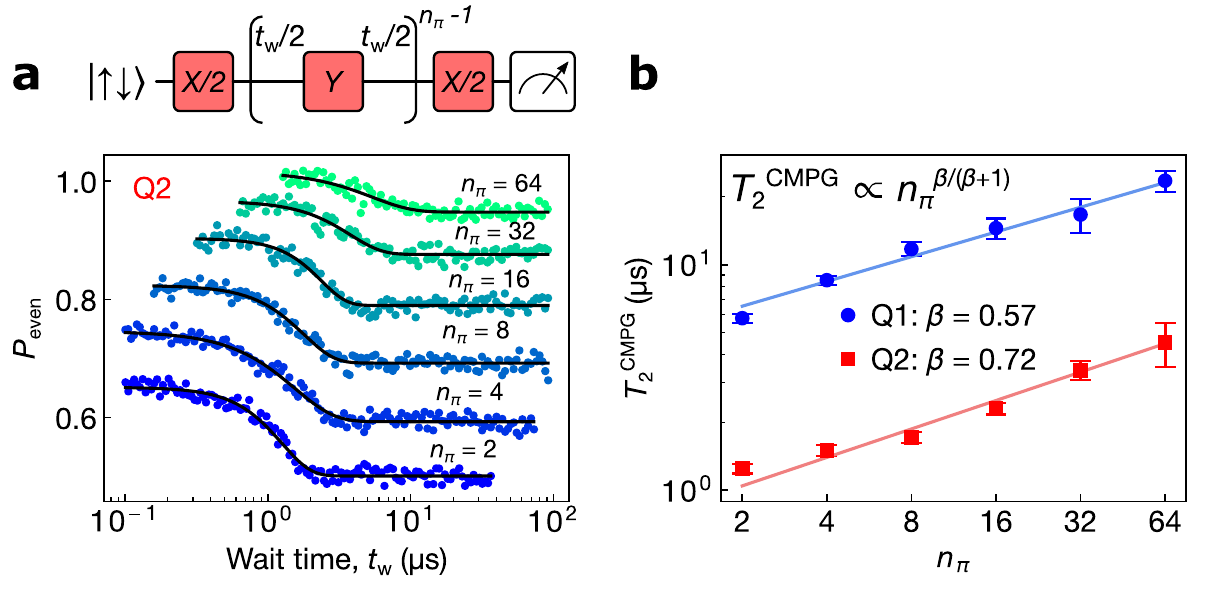}
    \caption{\textbf{CMPG and noise analysis.}
    \textbf{(a)} CPMG dynamical decoupling decays for Q2. The total number of $\pi$-pulses, $n_\pi$, is indicated near the respective decay trace. The decays are fitted to $T_2^\text{CPMG} = c_1 n_\pi^{\beta/(\beta+1)} + c_2$.
    \textbf{(b)} $T_2^\text{CPMG}$ time versus total number of $\pi$-pulses $n_\pi$, yielding $\beta = 0.57\pm0.13$ and $0.72\pm0.17$ for Q1 and Q2 respectively.
    }
    \label{fig:cpmg_analysis}
\end{figure*}

More generally, the echo experiment can be extended using the CPMG technique. By increasing the number of refocussing pulses, we expect an increase in the coherence times as the qubit is less exposed to 1/f noise. This is shown for Q2 in Supplementary Fig.~\ref{fig:cpmg_analysis}a, where we plot the resulting decay for different $n_\pi$.  In Supplementary Fig.~\ref{fig:cpmg_analysis}b, we plot the fitted $T_2^\text{CPMG}$ time as a function of the number of refocusing pulses $n_\pi$ for Q1 and Q2. By fitting $T_2^\text{CPMG} = c_1n_\pi^{\beta/(\beta+1)}$, we extract $\beta$ which represents the scaling exponent of a power-law noise spectrum, namely $S_\varepsilon(f) \propto f^{-\beta}$. These fits yield $\beta = 0.57\pm0.13$ and $0.72\pm0.17$ for Q1 and Q2 respectively. Furthermore, we compare these values of $\beta$ from the Hahn and CPMG decays through the relation $\alpha=\beta+1$. The decays yield an average of $\alpha=1.68\pm0.41$ and $1.79\pm0.29$ for Q1 and Q2, which corroborate the values of $\beta$ above.

\clearpage

\subsection{Spin relaxation time}\label{supplementary:T1_measurement}

\begin{figure*}[!ht] 
    \centering
    \includegraphics[width=0.5\linewidth]{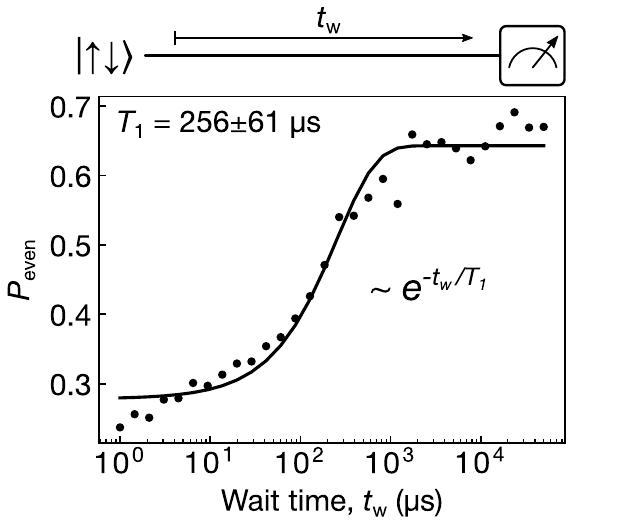}
    \caption{\textbf{Spin relaxation time.} We probe the spin relaxation through a $T_1$ measurement. Here, we initialise the state into $\UD$ before waiting for time $t_\mathrm{w}$ before performing readout. During the wait, the state will inevitably relax into $\DD$. We plot the resulting spin probability as a function of $t_\mathrm{w}$. After the spin relaxes, the resultant state maps to a even-parity outcome. The $T_1$ decay is described by the formula $c_1\exp(-t_\mathrm{w}/T_1)+c_2$, from which we extract the spin relaxation time $T_1 = 255\pm\SI{61}{\micro\second}$ measured at $B_\mathrm{ext}=\SI{0.83}{\tesla}$. This result compares similarly to the results measured in holes in silicon nanowires\cite{piot2023qubit}, and is on the same order of magnitude as recent theoretical predictions for this system \cite{wang2024electrical}. Additionally, the $T_1$ time in Ge has been observed to have a strong dependence on the size of the lead tunnel rates~\cite{hendrickx2020fast}. To this end, we note that the lead tunnel rates in our device were not optimised for $T_1$ time. Broadly speaking, these results reflect a common trend for many spin qubit experiments, where $T_1 \gg T_2$~\cite{stanoreview2022}, meaning that the limiting factor for these devices is not the spin relaxation time.
    }
    \label{fig:T1_decay}
\end{figure*}

\clearpage


\twocolumngrid
\clearpage
\bibliographystyle{naturemag}
\bibliography{Bibliographies/Full_arXiv}

\end{document}